\documentclass{svmult} 
\begin{document}

{\bf LIKELIHOOD   ESTIMATION OF  GAMMA  RAY  BURSTS
DURATION DISTRIBUTION}

\centerline{\it Istvan Horv\'ath }

\centerline{ Bolyai Military University, Budapest,  \ 
e-mail: hoi@bjkmf.hu }

Two classes of Gamma Ray Bursts have been identified so far, characterized
by $T_{90}$ durations shorter and longer than approximately 2 seconds. 
It was shown that the BATSE 3B data allow a good fit with three 
Gaussian distributions in $\log T_{90}$ \cite{HO}. 
In the same Volume in ApJ. another paper suggested that the third 
class of GRBs is may exist \cite{MUK}.
Using the full BATSE catalog here we present the maximum likelihood
estimation, which gives us 0,5\% probability to having only two subclasses.
The MC simulation confirms this probability.

\

In the BATSE current catalog \cite{M6} there are 2702 Gamma-Ray 
Bursts (GRBs), of which 2041 have duration information. 
\cite{K1} have identified two types of GRB based on durations, 
for which the value of $T_{90}$ (the time during which 90\% of the 
fluence is accumulated) is respectively smaller or larger than 2 s. 
This bimodal distribution has been further quantified 
in other papers \cite{K}, \cite{Kos}
 where a two-Gaussian fit were made.
Previously we have published an article \cite{HO}, 
where two and three Gaussian
fits were made using the $\chi  ^2$ method, which gave us
app. 0,02\% significance the third group is needed.
This is an agreement with the  \cite{MUK} result, who used a
multivariate analysis and find that the probability of existence of two 
clusters rather than three is less than 10$^{-4}$.
\cite{HAK1} also confirmed this result by statistical
clustering analysis, however they suggested
the third group was caused by instrumental biases
\cite{HAK1}, \cite{HAK2}.
Recently, remarkable anisotropy was found in
the angular distribution of this third group \cite{MBHBV}.
In this paper we take another
attempt at the trimodal distribution, evaluating the 
probability that the two populations are
independent using the maximum likelihood estimation.

For this investigation we have used a smaller set of 1929 burst 
durations in 
the current catalog, because these have peak flux information as well.
Firstly we take a two Gaussian fit for the duration which gives us a
best parameters of the two Gaussian fit,
which are very similar than previously was published \cite{HO}.
Secondly we take a three Gaussian fit. 
The means are -.25; .63; 1.55 in lg{\it s}.
These fits gives us the best logarithm's of the likelihoods
12320.11 and  12326.25.
Twice of the difference of these 
numbers follows the $\chi  ^2$ distribution with three degree of
freedom because the new fit has three more parameters \cite{P}.
The difference is 6.14 which gives us a   0.5\%
probability.
Therefore the third Gaussian fit is much better and
 there is a 0.005 chance the third Gaussian is caused by
statistical fluctuation.

One can check the probability using the Monte-Carlo (MC) simulation.
Generate 1929 numbers for T$_{90}$
whose distribution follow the sum of two
  Gaussian distributions.
Find the best likelihood with five free parameters
(two means two sigmas and two weights, but the sum of the last
two must be 1929). Secondly made a fit with three Gaussian
(eight free parameters, three means, sigmas and weights).
Take a difference between the two logarithm's of the
maximum likelihoods, which gives one number.
We do the process 100 times and have a hundred MC simulated
numbers. Only one of these numbers is bigger than
which the BATSE data has (6.14). 
Therefore the MC simulation confirm the mathematical
low statement and gives us a similar probability if
the third group is a statistical fluctuation.

The BATSE on-board software tests for the existence of bursts by comparing 
the count rates to the threshold levels for  three separate time intervals: 
64, 256, 1024 ms. The efficiency changes in the region of the middle 
area because the 1024 ms trigger is becoming less sensitive as burst 
durations fall below about one second. This means that at the ``intermediate"
timescale a large systematic deviation is possible. 
To reduce the effects of trigger systematics in this region we 
truncated the dataset to include only GRBs that would have triggered BATSE 
on the 64 ms timescale.
Using the Current BATSE catalog CmaxCmin table \cite{M} we choose the GRBs,
which numbers larger than  one in the second column (64 ms scale
maximum counts divided by the threshold count rate).
Although this process reduced the bursts numbers very much  
(only 857 GRBs remain)
the significance level still stay below 1\%.

It is possible that the three log-normal fit is accidental, and that there 
are only two types of GRB. However, if the T$_{90}$ distribution of 
these two types of GRBs is log-normal, then the probability that the third 
group of GRBs is an accidental fluctuation is less than 0.5-1.0 \%.

This research was supported in part through OTKA F029461 and T34549. 
Useful discussions with M. Briggs,
E. Fenimore, J. Hakkila, P. M\'esz\'aros,  are appreciated.

\end{document}